# Bluetooth Smartphone Apps: Are they the most private and effective solution for COVID-19 contact tracing?

**Scott McLachlan**[1,2], **Peter Lucas**[3], **Kudakwashe Dube**[2,4], **Graham A Hitman**[5], **Magda Osman**[6], **Evangelia Kyrimi**[1], **Martin Neil**[1], **Norman E Fenton**[1]

[1] Risk and Information Management, Queen Mary University of London, United Kingdom
[2] Health informatics and Knowledge Engineering Research (HiKER) Group
[3] Faculty of EEMCS, University of Twente, Netherlands
[4] School of Fundamental Sciences, Massey University, New Zealand
[5] Centre for Genomics and Child Health, Blizard Institute, Queen Mary University of London, United Kingdom
[6] Biological and Experimental Psychology Group, Queen Mary University of London, United Kingdom

**Abstract**

Many digital solutions mainly involving Bluetooth technology are being proposed for Contact Tracing Apps (CTA) to reduce the spread of COVID-19. Concerns have been raised regarding privacy, consent, uptake required in a given population, and the degree to which use of CTAs can impact individual behaviours. However, very few groups have taken a holistic approach and presented a combined solution. None has presented their CTA in such a way as to ensure that even the most suggestible member of our community does not become complacent and assume that CTA operates as an invisible shield, making us and our families impenetrable or immune to the disease. We propose to build on some of the digital solutions already under development that, with addition of a Bayesian model that predicts likelihood for infection supplemented by traditional symptom and contact tracing, that can enable us to reach 90% of a population. When combined with an effective communication strategy and social distancing, we believe solutions like the one proposed here can have a very beneficial effect on containing the spread of this pandemic.

# Introduction

At the time of writing many of us are in our fifth or sixth week of social distancing and lockdown in an effort, we were told, that would *flatten the curve* and curtail the spread of COVID-19. As considerations move from dealing with the worst of the disease to containment of any remaining pockets of infection, much noise is being made in the media concerning the need to implement *contact tracing apps* (CTA) before the world can return ostensibly to normal (Mathews, 2020; Scott, 2020; Whittaker, 2020; Drew 2020). While the claimed benefits for CTA of being able to leave our homes, reopen workplaces and revive crippled economies are significant, CTA are not without some controversy (Lomas, 2020; Volk, 2020). Questions regarding transmission dynamics and optimal intervention strategies for the disease, and the risk CTA pose to individual privacy and efficacy are repeatedly raised, and many feel these have not been adequately answered (Crocker et al, 2020; Sun & Viboud, 2020). Some describe CTA as the *trojan horse*: reminding us that many governments and corporations already operate population-wide electronic surveillance and the likelihood that they do not, and once they also get access to our CTA data, will not, act in good faith (Lomas, 2020). However, what everyone fails to ask is whether this personal information is being provided in support of the most, or even an effective method and at what uptake rate in the general population are we sure that it will be worthwhile. Is a Bluetooth radio beacon paired to a smartphone app the most effective method for digital contact tracing? In this paper we address these key questions for smartphone-based contact tracing solutions.

## What is *Contact Tracing*?

Proposed more than 80 years ago for the control of syphilis (Paran, 1937), contact tracing is a surveillance and containment strategy for infectious disease (Vazquez-Prokopec et al, 2017). Rather than managing only isolated cases as they seek medical attention, contact tracing follows the path of infection from diagnosed patients to those with whom they have been in close physical contact (Armbruster & Brandeau, 2007; Eames, 2007; Vazquez-Prokopec et al, 2017). Several approaches for contact tracing have been described in the literature, including: first-order, single-step, iterative and retrospective (Eames, 2007; Klinkenberg et al, 2006). *First-order* tracing only identifies those people the patient immediately came into contact with, and advises them of potential exposure and the need to seek medical advice or self-isolate. It does not concern itself with tracing the contacts of contacts, leaving that second-order process to occur as and when the first-order contact seeks medical care. *Single-step* contact tracing identifies all people that the infected person came into contact with, and as any of those are also identified as infected, their contacts are identified and the process continues. One issue with single-step contact tracing is that asymptomatic infecteds can spread the disease until they are detected and isolated. In contrast, *Iterative* contact tracing continues to track and re-apply the relevant diagnostic test to contacts iteratively before their infection may even be detected through symptom screening. The process continues until no further infecteds are identified. The final type, *Retrospective* contact tracing, follows the same process as either single-step or iterative with the addition that it also operates in reverse by considering the people with which the infected patient had been in contact with in their recent past, with the goal to identify who it was that infected the patient. Each approach is demonstrated in Figure 1.

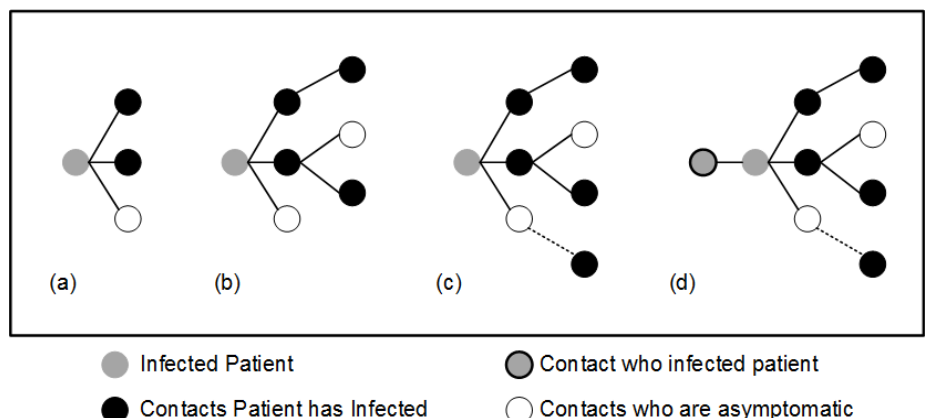

**Figure 1:** (a) first order; (b) single-step; (c) iterative; and (d) retrospective contact tracing.

Contact tracing has traditionally been conducted as a manual multi-stage process that begins when a patient is diagnosed with an infection that is usually also subject to notification rules that require the clinician to apprise the health authority (HA) of the infected's status. Any likely contacts of the infected patient are determined, identified, advised of their exposure status and encouraged to seek medical advice (Armbruster & Brandeau, 2007; Eames, 2007). Generally, contact tracing has only been used for diseases with *low prevalence*: meaning diseases where there is only a small number of cases in the community at any given time (Armbruster & Brandeau, 2007). Examples of diseases where contact tracing has been applied include: tuberculosis, HIV/AIDS, Ebola and sexually transmitted diseases (Armbruster & Brandeau, 2007; Danquah et al, 2019; Eames, 2007; Yasaka et al, 2020). On review, many of these examples show the efficacy and reliability of contact tracing to be uncertain and contentious issues.

## Modern contact tracing using wireless beacons

With our vastly increased global population, international airline travel, megacities and mass transit, it is unlikely that traditional contact tracing alone could contain even a minimally contagious disease (Niehus et al., 2020). Traditional contact tracing was used early-on during the SARS epidemic (Fidler, 2004; Huat, 2006). However, it failed to contain the infection which quickly spread through the wider community, with global HAs realising that new approaches were now required (Fidler, 2004; Huat, 2006). Modern contact tracing approaches have been proposed using ubiquitous and pervasive smartphones and the wireless technologies they contain to record and report when we have come into close physical contact with others. It is believed this automated contact tracing will overcome situations when we either are not aware of, or don't recall, every contact incident (Maghdid & Ghafoor, 2020). The proposed approaches shown in Figure 2 incorporate these technologies to more efficiently and effectively provide: (a) movement-focused mobile-assisted automatic contact recording; (b) contact identification; (c) contact notification; and, (d) narrowcast messaging (Maghdid & Ghafoor, 2020; Vazquez-Prokopec et al, 2017; Yasaka et al, 2020). Proponents of CTA claim, possibly disingenuously given their extensive and publicly-funded investment in development of the app, that installing the app will significantly reduce the chance of you passing on the infection to your family and friends (COVIDSafe App, 2020), and essential to keeping your family safe from COVID-19 (Hamilton, 2020).

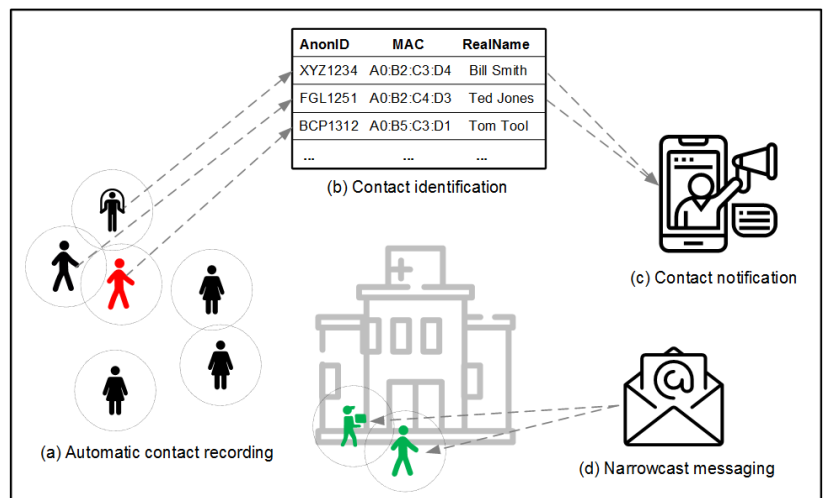

**Figure 2:** Modern applications of contact tracing using smart devices

(a) CTA automatically records anonymous IDs of other devices that come within the broadcast distance of the wireless technology being used (Bluetooth or wifi).
(b) Central server operated by health authority or a technology supplier maintains the linking table that can identify all users of CTA.
(c) When an infected person (in RED) notifies the CTA of their positive diagnosis, the central server advises all CTA users who have been in close physical contact with the infected that they should seek medical advice.
(d) The central server can also be used to send narrowcast messages, for example: alerting people who CTA location tracing identified near a particular infection hotspot during a defined period (in GREEN) that they may have been exposed and to seek medical advice.

While solutions using WiFi MAC address sniffing (Lu et al, 2020), GPS (Finazzi, 2020; Klopfenstein et al, 2020; Maghdid & Ghafoor, 2020) and cellular network geolocating (DP3T, 2020; PEPP-PT, 2020) have all been proposed, many believe Bluetooth tracing to be the most suitable for use in CTA (Berke et al, 2020; Brack et al, 2020). Authors point to the fact that Bluetooth has already been demonstrated effective for proximity detection (Berke et al, 2020; Brack et al, 2020). It is also claimed that while Bluetooth has an effective range of around 25-30 metres, signal strength can be used to effectively identify whether another device is within the 2-metre rule promoted as a component of social distancing (Berke et al, 2020; Brack et al, 2020; Xia & Lee, 2020).

# CTA Data Points and Privacy

Most attention to privacy in the literature focuses on the interactions and data passing between users of the CTA when they come into close physical contact and their devices *handshake*. A smaller focus is given to interactions between the CTA and HA server, whose privacy exposure is mitigated, it is claimed, by *decentralised* solutions: that is, solutions where most data remains on the user's device and only small push or pull transactions occur to the HA server to either advise the system of the user's COVID-19 diagnosis, or verify that the user has not already been in contact with another who has since been diagnosed. What is clear is that while labelling their solutions as *privacy-preserving*, most authors seek to mitigate one form of data or privacy loss while ignoring, intentionally or not, every other possible disclosure vector (Kuhn et al, 2020). To the best of our knowledge, no author considered the issue of metadata and its effect in nullifying their often complicated and expensive privacy solutions.

Metadata is the most common and easily accessible form of personal information being collected (McLachlan, 2016). Metadata is defined as information *about* a communication: the *who*, *when*, *where*, and *how* but not the *what*. Metadata contains sufficient information to know when you made a call, texted, emailed or accessed a web page, who your communication or web request was made to, how and whether the person or system at the other end received the communication. The only thing metadata does not contain is the actual content of the message (Maurushat et al, 2015). For more than a decade metadata has been used by law enforcement and others to draw inferences about our state of mind, intentions, previous travel, personal associations and interactions (Maurushat et al, 2015; McLachlan, 2016). In many countries metadata may be accessed without a warrant by authorised organisations and agents, and laws exist requiring telecommunications, internet service provider companies and web hosts to maintain large stores of metadata collected as a result of the activities of individual subscribers (Maurushat et al, 2015; McLachlan, 2016; Shamsi et al, 2018).

Let us consider the data that is generated while using a CTA. Figure 3 presents the typical CTA use-case described by many authors, in which: (a) the primary CTA user and others install and register the app on their smartphones; (b) as they move around and come into close physical contact with each other, their smartphones identify other smartphones and a contact trace is recorded; (c) an upload of some information passes from the CTA on the users device, via their provider's core network (cellular or ISP); (d) from their provider, via the internet, to the HA servers; and (e) alerts and updates can also be sent from the HA server to individuals, or every user. Some variation is observed in the literature claiming to present privacy-preserving methods regarding: (i) the type of information passed from the CTA to the health authority server; and (ii) whether the data passes directly to the HA server or, as with the Singapore (TraceTogether), Australian (COVIDSafe) and proposed Apple/Google collaboration examples, into a third-party supplier's international datacentre cloud network (i.e. Google, Apple or Amazon Web Services) before being received by the HA server (Maddocks, 2020).

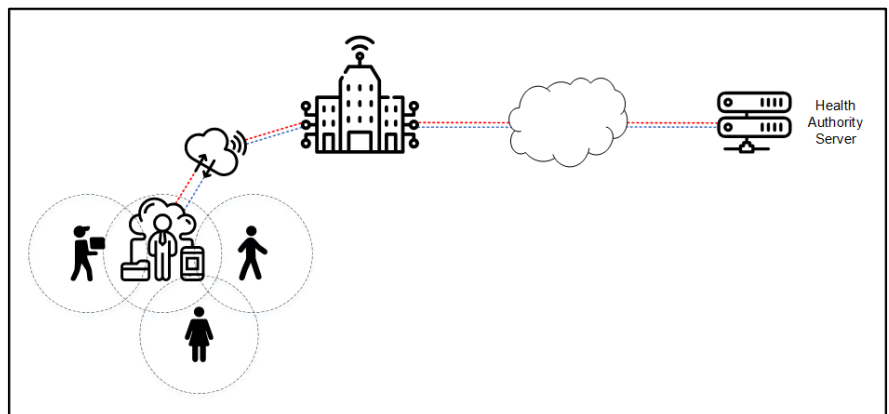

**Figure 3:** The Typical CTA scenario

Metadata are generated at every step of the typical CTA scenario. Every communication or request sent to cellular, internet service provider or web host organisations results in metadata that must be stored in logs in their network that identify you from your subscriber identity module (SIM) record matched to the details of your device, with a record of what you requested or sent, to or from whom, and when (De Carli et al, 2020; McLachlan, 2016; Shamsi et al, 2018). All digital traffic passing from your provider's network via the internet to the HA results in metadata being captured inin the systems of every network provider between the two, but more importantly, in the HA's network systems and servers. Believed by many to be *non-sensitive*, metadata often remains overlooked in smartphone and internet-facing solutions even though it can be a trivial matter to re-identify an individual and their actions and interactions with others from the metadata, or digital breadcrumbs, they create (Ho et al, 2018; Maurushat et al, 2015; Perez et al, 2018; Shamsi et al, 2018).

# CTAs for COVID-19

While Singapore and Australia's Health Departments have already commenced rollout of CTA solutions for COVID-19, the United Kingdom (UK), North America and most of Europe will only commence their trial deployments in the coming week (Hern & Sabbagh, 2020). Taiwan, South Korea and Israel were even more proactive, with increased testing, quarantines and mandated CTA of recent travellers and the infected that has resulted in lower rates of secondary infections and significantly fewer deaths, with alarms being raised, similar to home detention systems for criminals, informing police if those in quarantine left the building in which they were being housed (Lee, 2020; Lomas, 2020b). Most literature proposing CTA and being used by academics and governments to support efforts, efficacy and expenditure of public funds for COVID-19 contact tracing with smartphones, are theoretical solutions in hurriedly prepared preprints that are yet to undergo rigorous testing or peer review. Examples include: (Berke et al, 2020; Brack et al, 2020; De Carli et al, 2020; Hekmati et al, 2020; Klopfenstein et al, 2020; Maghdid & Ghafoor, 2020; Reichart et al, 2020; Xia et al, 2020). While acknowledging that privacy is not a design goal for any CTA, many propose solutions that they claim are *privacy-preserving*: both between app users generally, and between individuals and the health authority and technology suppliers who maintain the central servers (Brack et al, 2020; Reichert et al, 2020). Only one paper was identified in this work that acknowledged no privacy could exist where there was a central authority, and that users should only expect solutions to keep them blinded from each other (Berke et al, 2020).

Some solutions present as a confusing array of seemingly random technology, thrust together (Reichart et al, 2020). Apps proposing ID hashing or public/private key encryption between central server and end-user claim these additions ensure complete user privacy: and while authors acknowledge that the central server will have recorded your current and all previous hashIDs and will be used to distribute alerts to other users, they also disingenuously claim that the health authority are entirely unable to learn anything at all about users, the infected, or their contact history from this vast collection of data (Brack et al, 2020). Many proclaim CTA ineffective because it relies on willing individuals who must provide identifying information about themselves and those they come into contact with, and self-report their infected status via the app for storage on a central server (Brack et al, 2020; Hekmati, 2020; Yasaka et al, 2020). Usually, while simultaneously claiming to provide a decentralised or privacy-protecting solution that still uses user IDs and other information such as location or contact lists that are uploaded or shared via the central server (Brack et al, 2020; Hekmati et al, 2020; Reichert et al, 2020; Yasaka et al, 2020). However, decentralisation adds complexity (Berke et al, 2020), often without a significant improvement in privacy. In one case an infected still self-reports, except that they are instead required to provide a signed medical certificate, exposing even more personal information to whomever runs the central server so that an alert can be broadcast to others who the infected has previously been in contact with (Hekmati et al, 2020).

# Efficacy of CTA

Researchers and epidemiologists have sought, somewhat unsuccessfully, to understand the efficacy and overall value of disease contact tracing for many years, with heightened interest often observed in the aftermath of disease outbreaks. Many issues limit contact tracing efficacy, the most significant being the need to understand transmission, susceptibility, prevalence, and latency for the target disease (Kiss et al, 2005). Before deciding on an effective control strategy, it is essential to understand the course of the disease. In epidemiology, many compartmental models have been developed for modelling infectious diseases (Roddam, 2001; Hethcote, 2000). One commonly used model computes the theoretical number of people infected with a contagious disease in a closed population over time is the *Susceptible-Infected-Recovered* (SIR) model (Anderson, 1991; Rodrigues, 2016). These mathematical models are being considered an important source of knowledge for global governments making life-or-death decisions regarding management of COVID-19. The *Susceptible-Exposed-Infected-Recovered* (SEIR) model has been used to focus on transmission of COVID-19 in Wuhan, China (Lin et al, 2020), and to compare outcomes for different containment policies (Casella, 2020). The *Susceptible-Infectious-Recovered-Dead* (SIRD) model has been used to provide estimations of the basic reproduction number (R0), per day infection mortality and recovery rates, and attempts to forecast the evolution of an outbreak at the epicentre three weeks in advance (Anastassopoulou et al., 2020). *Susceptible-Infected-Diagnosed-Ailing-Recognized- Threatened-Healed-Extinct* (SIDARTHE) was proposed as an extension to SIR in an effort to model the COVID-19 epidemic in Italy (Giordano et al., 2020). Their model showed that enforced lockdowns could be mitigated in the presence of widespread testing (Peto, 2020) and contact tracing, strongly contributing to rapid resolution of the epidemic. Similar findings were also found in (Hellewell et al., 2020). While some believed contact tracing was effective during the SARS outbreaks of the early 2000's (Kiss et al, 2005), we have already discussed Singapore's reliance on contact tracing during that period and how on review it was found to have failed (Fidler, 2004; Huat, 2006). Other examples where contact tracing failed, in some cases even with the use of smartphone technology and apps, include an audit of contact tracing use for tuberculosis (Hussain et al, 1992; Mwongela, 2018); the Foot and Mouth outbreak in the UK in 2001 (Kiss et al, 2005; Kao, 2003); and the 2014-16 Ebola epidemic in West Africa (Danquah et al, 2019). While many claim suitability, viability and effectiveness for CTA, in most cases the CTA solution they propose has yet to be prototyped, and for those that were, trialled in anything approaching a real-world situation (Brack et al, 2020; De Carli et al, 2020; Hekmati et al, 2020; Klopfenstein et al, 2020; Mwongela, 2018; Yasaka et al, 2020).

We sought to understand how effective CTA might be as a containment approach for COVID-19 in highly populous locations like London or Birmingham in the UK, or Sydney and Melbourne in Australia. We observed that most papers presenting a CTA appeared to silently apply best-case assumptions when discussing or evaluating their models in order to paint their solution in the best light. For consistency, we chose to continue this practice albeit with the novel addition of transparency. With respect to how many people an infected person may come into contact with, we rely on the calculations provided in the UK that have come to be known as *the Oxford figures* and have been used by those developing and promoting the need for an NHS-specific app, and in the media, to support efficacy, funding and deployment of the NHS app (Merrick, 2020). The authors used an SEIR model to suggest that in a 14-day period post-lockdown the average person comes into contact with 217 people, of which 59 are considered to be close contacts sufficient for disease transmission, and of those 36 would be individuals in a CTA scenario who are potentially traceable (Keeling et al, 2020). While we could have worked from the number of 59 close contacts which would have made our numbers significantly larger and more dramatic, in order to demonstrate the fallacy of claims made in support of CTA even as a component in disease containment for COVID-19, we chose again to work from a best-case position and elected to use latter and much lower figure for total transmissions. The Oxford figures also provide that the average latent period, usually defined as the period between when a person is exposed to the virus and when they begin exhibiting symptoms, is 4 days (Keeling et al, 2020). Other authors using larger datasets provided this incubation period was 5 days, with 97% of patients showing symptoms at day 12 (Lauer et al, 2020; Qi et al, 2020). Younger infected patients tend to be asymptomatic, and for longer periods, and while the mean serial interval, the time between when symptoms appear in infector and infectee) varies between 4 and 7.5 days (Du et al, 2020; Qi et al, 2020). It should be noted that our best-case assumptions are similar to those of Dr Hannah Fry's group (Kucharski et al, 2020) except that our mean delay from symptoms to isolation was reduced to 1 day: the effect of which would be to reduce the number of secondary infecteds created by each primary in our scenarios. In spite of this, our results were statistically similar to those of Kucharski et al (2020).

> **Literature Review**
> We searched PubMed, medRxiv, bioRxiv, arXiv and DOAJ for peer-reviewed articles and preprints that mentioned the terms "contact tracing", and "COVID-19". Our initial search revealed more than 1300 articles published since December 2019. We narrowed our search to those articles published since March 2020 and selected only those that proposed a CTA solution. This identified a collection of 72 papers whose solutions were reviewed. From these papers 59 (89%) used the term privacy in either the title, abstract or introduction, and 52 (72%) proposed solutions claimed to be privacy-preserving. Solutions intended to reduce or eliminate data passing to a central server, described as decentralised solutions, were proposed in 19 (26%). Only 3 (4%) solutions described production of a prototype with 1 (1%) solution having been tested with simulated data.

The assumptions used in our calculations include that:
  a) The infection clock starts from exposure;
  b) From day 5 the infected begins to shed the virus;
  c) Patients may become symptomatic between days 5.5 and 11.5;
  d) At day 12 every infected is considered to by symptomatic;
  e) Each infected comes into close contact with 36 people in a 14-day period, pro rata for the period between day 5 and when they become symptomatic;
  f) Every infected has self-isolated from day 13;

For the 6 o'clock path shown in Figure 4, we present the absolute best-case scenario where 100% of the population have smartphones, install the CTA, are tested, immediately self-report and self-isolate. This scenario, whilst being quite impossible, would actually contain the disease in only two cycles, or 14 days.

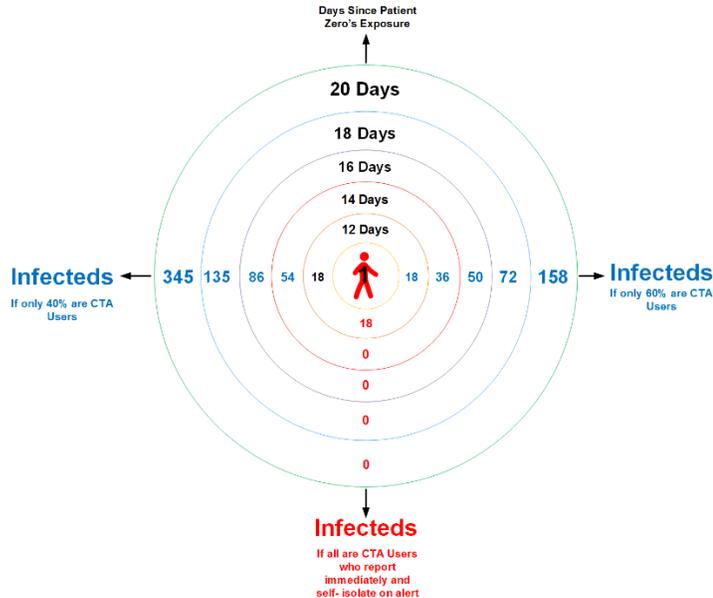

**Figure 4:** COVID-19 CTA Infection Scenarios

The 3 and 9 o'clock paths present the UK and Australian scenarios for the claimed 60% (Merrick, 2020) and 40% (Woodley, 2020) adoption that we are told would deliver CTA success in their respective populations. In each scenario every infected spreads COVID-19 to only a small number of infecteds, and while a percentage of secondary infecteds are alerted through the CTA and self-isolated, the remaining percentage, those without the app, persist to spread the infection to a significantly large number of people. Figure 5 provides a visual representation of the progress at each stage for the 60% adoption NHS CTA scenario.

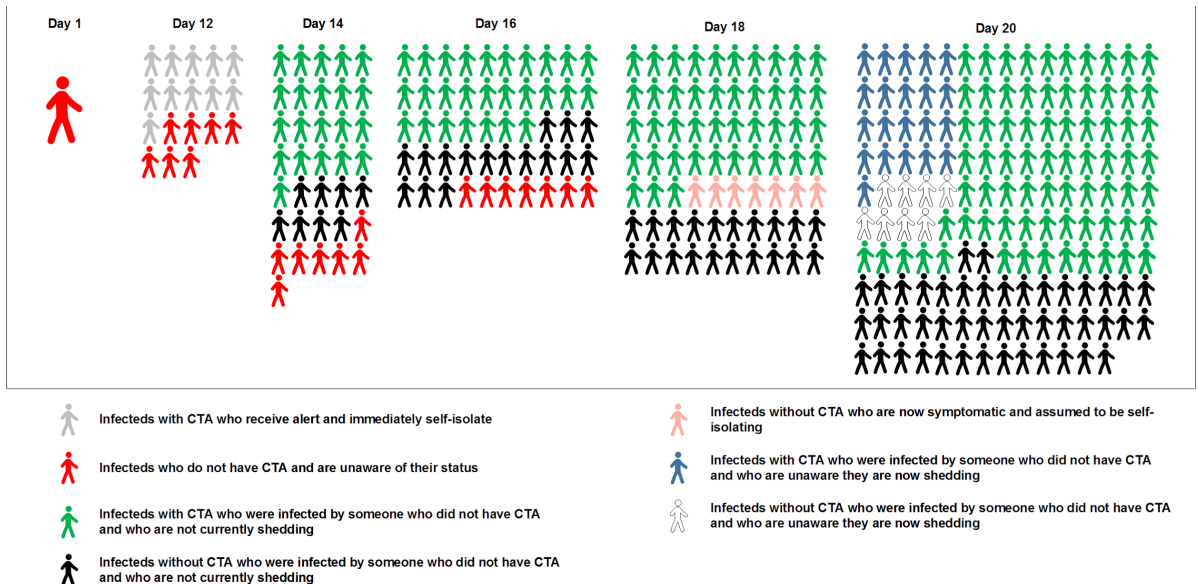

**Figure 5:** Visual representation for the 60% CTA user NHS scenario

Smartphone penetration for adults in the UK has only achieved 79%, reducing to 40% in the key COVID-19 demographic, the over-65s. Australian figures are similar. To get 60% penetration in the overall UK population, more than three quarters (76%) of all smartphone owners must install, register and use the app. This assumes absolutely no loss to follow-up, which occurs where a user either stops using or removes the app from their device for any reason. When the average loss to follow-up in a clinical trial is 6% (Akl et al, 2012), the NHS app would actually require more than 82% of the smartphone-owning population to initially install and register the app to increase the probability that 60% will use their CTA to completion.

For the 40% (Australian) and 60% (UK) scenarios we begin from the position that 40% and 60% of the population respectively have installed the app and immediately self-report and/or self-isolate when alerted. As these scenarios played out, we calculated under an absolute best-case wherein people who were alerted by the app or who reached day 13 all immediately self-isolated. The issue with this is that we know some people's symptoms will not be severe enough at first for them to believe they have the disease and seek medical advice. This is human nature. Studies report that around 18% of all exposed people remain asymptomatic but recover from the virus in a timeframe similar to that of people who do become symptomatic (Mizumoto et al, 2020; Day, 2020). Further, 1-2% of patients will be asymptomatic but remain contagious and continue to shed the virus from 1-3 months after their initial exposure, with or without a symptomatic period (Bengali, 2020). In keeping with our best-case model we have not incorporated additional potential exposures that would arise from these groups of people in our calculations.

A final set of calculations was performed seeking the sweet spot: that number below absolute for CTA adoption in the overall population where the number of secondary cases was manageable by manual contact tracing and other containment methods, and the NHS generally. Table 1 presents the results of those calculations and, similar to figures proposed by other groups who have evaluated this issue (Bulchandani et al, 2020), we find the sweet spot for CTA uptake in order to control COVID-19 lies somewhere between 90 and 95%. As discussed, such high uptake is simply not credible or possible.

**Table 1:** Number of additional infecteds created per one infected, based on % of the population installing and immediately complying with the CTA

| Day: | 12 | 14 | 16 | 18 | 20 |
|---|---|---|---|---|---|
| 95% | 18.0 | 4.5 | 4.7 | 5.1 | 6.2 |
| 90% | 18.0 | 9.0 | 9.9 | 11.3 | 16.0 |
| 80% | 18.0 | 18.0 | 21.6 | 27.0 | 46.8 |

*NB - While estimates suggest 94% of UK adults owns a mobile telephone (https://www.tigermobiles.com/blog/mobile-phone-usage-statistics/) only 79% of those over 18 in have a smartphone (source: https://www.finder.com/uk/mobile-internet-statistics ) and only 40% over 65 - the key demographic for infection and death from COVID-19.

# Proposed and current COVID-19 Solutions

This paper has considered many of the barriers that continue to impede success for contact tracing, even when it is automated with a smartphone app. We now turn to consider the current or proposed solutions and how, even if not completely successful, they might be better designed and promoted in order to produce a lasting benefit for the average individual and wider community.

## Solution Option 1: *Oxford/NHS App*

The United Kingdom breakout box describes the app being rolled out by the UK government and NHS. We refer to it as the Oxford/NHS app since its development was led by academics at Oxford. The government are pinning their hopes on this app being a key enabler for relaxing the current lockdown policy. Appendix 3 discusses the common properties and data being collected by CTA reviewed during this research. We believe the statistics and overall proposal to support development of the app and promote its uptake in the community are based on best-case scenarios. However, we do perceive that the strength of Government and NHS support comes from the perception of trust they seek to engender. The openness and degree of transparency that the NHS and Oxford teams have been upselling in the media, if delivered, far exceed that of any other. We found no other State-developed or operated solution that suggested a willingness to allow the media, technologists and general public access to the source code. However, early non-published results of a pilot trial on the Isle of Wight are less encouraging with a major limiting factor being the variation in smartphone operating systems, especially those of older phones (Duell, 2020). The level of transparency underpinning the NHS solution needs to also be adopted in any use of the APIs provided by the Apple/Google collaboration.

## Solution Option 2: Chan/Spector Symptom Tracker App merged with CTA

In late March 2020 a collaboration between Massachusetts General Hospital, Harvard Medical School, Guys and St Thomas' NHS Foundation Trust and King's College London released their COVID-19 Symptom Tracker (Drew et al, 2020). They currently have 2.8 million users and report symptoms data gathered from around 1.6 million, of whom only a tiny fraction of 1,176 (0.07%) had undergone some form of PCR-based diagnostic test (Drew et al, 2020). Many issues may present with this type of study. These issues include the subjective nature of the endeavor, the bias that comes from the fact that the app was initially promoted to and installed by clinical staff and their families, and that many in the wider community who voluntarily install such apps are the worried well who, when prompted with questions suggesting the symptoms that go with a condition, are more likely to identify as having some of them. Unless carefully managed, suggestibility unintentionally induces conditioned associations between symptoms, leading patients to report more intense or additional flu-like symptoms (Skelton et al, 1993).

Leaving these issues aside we believe a good solution might have been to incorporate CTA into this Symptom Tracker app, and allow the existing user-base to either consent or decline providing that additional information. That a high number of existing users would consent to the addition is far more likely than believing that almost 3 million people will install a second COVID-19 related app. We also believe that any proposed CTA solution should contemplate capture of many of the same symptom-based data-points, whether used directly in contact tracing or not. We suggest this in order to enable future anonymous aggregation and data mining/knowledge engineering on COVID-19 from what could be a considerably much larger and richer dataset.

## Solution Option 3: Bayesian network based COVID-19 CTA

Our proposed solution focuses on enabling users to diagnose the possible presence of Covid-19 themselves. This is done through a causal probabilistic model (a Bayesian network, that we describe in Section 4.1) that is made available in a smartphone app based on the architectural framework (that we describe in Section 4.2). The app provides the user with information about how likely it is they have or have not mild or severe Covid-19. When this probabilistic information is combined with data about the GPS-location of the smartphone, together with information about the age group of the person the triple *(Prob. user has Covid-19, GPS-location, Age-group)* can be used to provide information about the distribution of mild and severe Covid-19. For example using colour shades on the map of a country, the data can be used to present a dynamic visualization of the probability distribution on the location where that information was collected (Hay et al, 2013). *This solution option involves providing diagnostic-oriented feedback to citizens with real time Covid-19 surveillance and minimal privacy infringement as quickly as possible in the face of all the limitations of the current constantly changing situation.* Response measures from the information collected from this option operate mainly at the population location level, such as intensified lockdown, social/physical distancing and self-isolation campaigns rather than more granular contact tracing and individual isolation measures requiring massive resource deployment. This option is dramatically different from the many trace and contact app solutions provided elsewhere.

## 4.1 The Bayesian network (BN) for providing feedback on user symptoms

A Bayesian network (BN) (Cowell et al, 1999; Fenton and Neil 2018, Koller & Friedman, 2009; Pearl, 1988) is a graphical model consisting of nodes and arcs as shown in Figure 4 (this is the draft model we propose for our app). Some of the variables (such as those representing symptom nodes) may be directly observable while others (such as the COVID-19 node) are not. There is an arc between two nodes if the corresponding variables are causally linked in a probabilistic sense. The strength of the link, as well as the uncertainty associated with these, is captured using probabilities and statistical distributions. When data are entered into the model for specific variables that are observed, all of the probabilities for, as yet, unknown variables are updated using an AI algorithm called Bayesian inference. Hence, in the model here, the BN algorithm computes the probability of having none, mild, or severe COVID-19 , based on present *signs and symptoms* and other relevant background information entered by the user.

The model makes a number of simplifying, but rational assumptions. For example, it assumes: that a person can only become infected if they have been in recent contact with an infected person (or some biological matter from an infected person); that a positive test result from a perfectly accurate COVID19 test procedure would mean that the person has COVID19 (even if they were asymptomatic); that there may be other conditions such as COPD or flu that have some symptoms in common with COVID-19.

The probability distributions in the model for the symptoms given the disease status (i.e. the status of the COVID-19 variable) are based on the statistics provided in the paper by Huang et al. (2020). All the assumptions are described in Appendix 5.

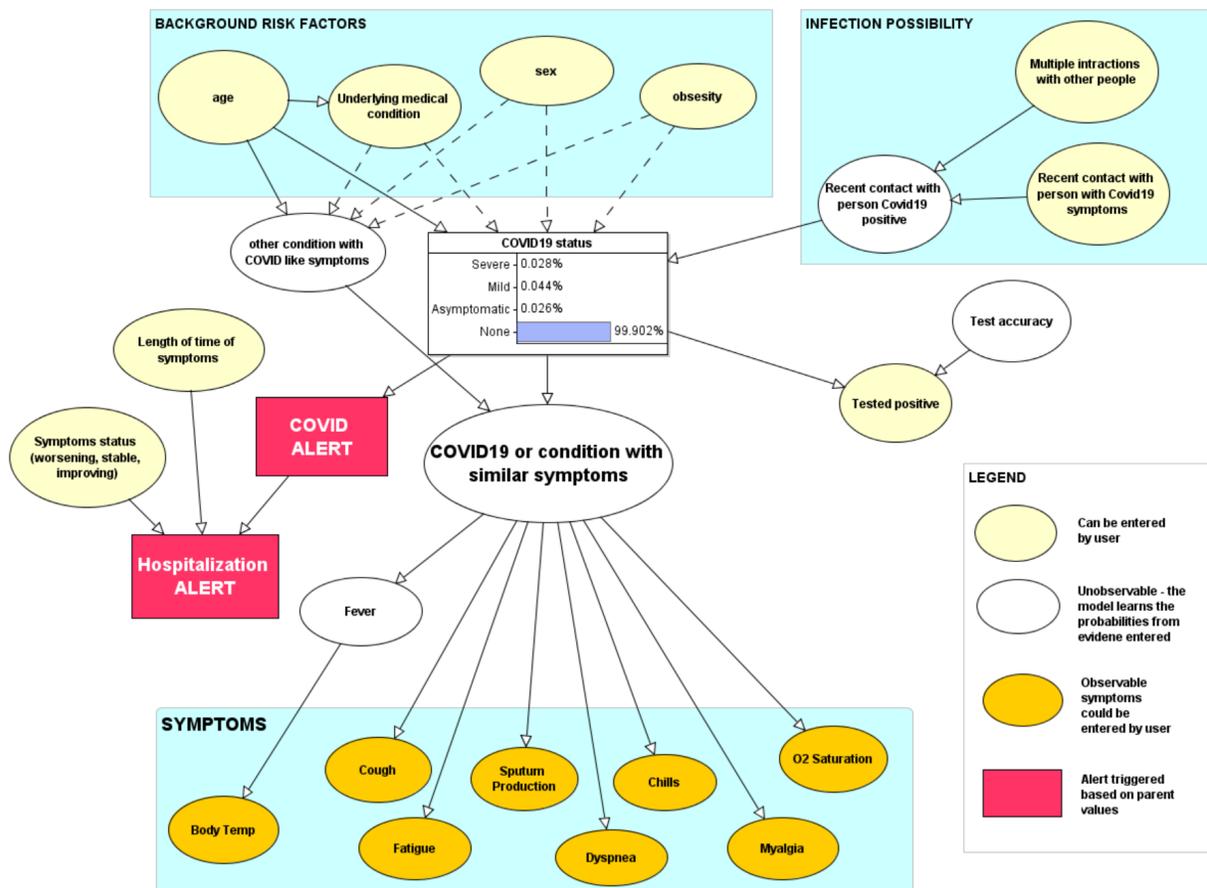

**Figure 4:** Covid-19 Bayesian network model structure. The probabilities shown for the COVID19 status node represent the prior probabilities when no observations are entered.

Figure 5 shows the updated predicted probabilities with some user entered observations; in this example a user has many of the COVID19 symptoms and has had multiple recent interactions with other people. Although this user has not entered their background or risk factors, the model infers there is a 76% probability the person has Covid19 (66% probability severe and 11% probability mild). Note that the model also updates the probabilities for the unknown risk factors and background nodes. For example, this person is more likely to be male than female (56%) and is likely to be over 65 (54% probability). The probability of obesity is 12% (up from a prior of 10%). These backward inferences are simply the application of Bayes. Appendix 5 illustrates the power of the model through other scenarios.

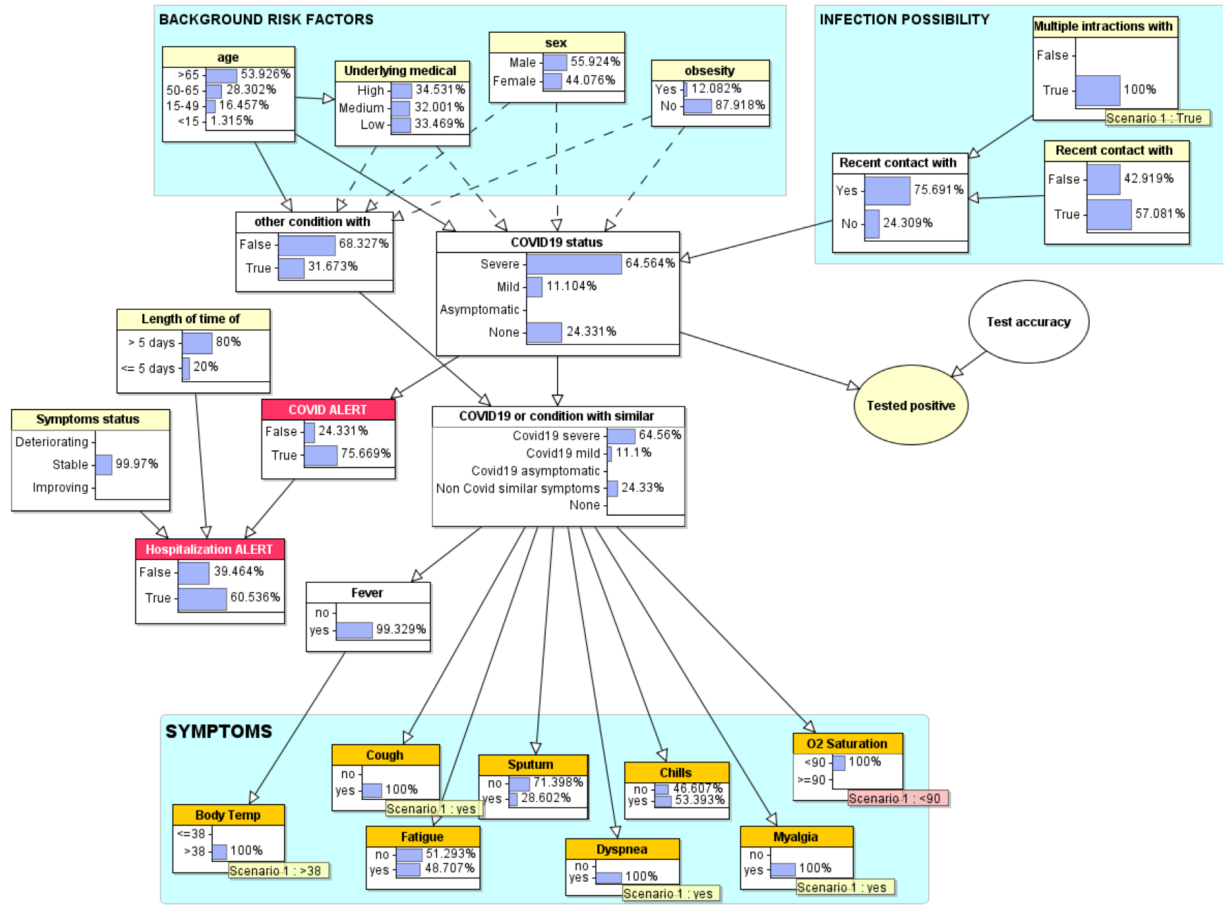

**Figure 5:** Covid-19 BN model for a user with most CODID19 symptoms and multiple recent interactions with other people (nodes with observations are denoted with a scenario label).

Depending on the value of the 'alert threshold' that is set the model will trigger an alert (it will trigger a separate hospitalization alert depending on the length of time the symptoms have been present and whether or not they are improving). So those people with the app who have come into contact with the person will be alerted that they have been in contact with a person most likely to be Covid19 positive, while this person could be given appropriate instructions for contacting the health authorities.

This model is still an incomplete attempt at developing a BN for the prediction of the presence of Covid-19 (we are in the process of gathering the relevant data required to complete all of the probability tables; currently those for which we do not have relevant data, or are not logically determined, are simply estimated). It is possible to add other signs and symptoms (for example dizziness seems useful) and also comorbidities and immunodeficiency could be added, as the literature provides the relevant information.

The advantage of a BN is that it can still generate predictions with incomplete information. Thus, if certain evidence is not entered by the user, the model is able to use prior probabilistic information rather than make particular assumptions. So, although *body temperature* and *oxygen saturation* are key measurements, the user decides whether or not these measurements are actually done. Using the BN it is also possible to predict which feature will be the most informative one in contributing to the diagnosis, and this feature can be used to request additional information from the app's user after some initial input.

## 4.2 Design of the BayesCOVID Surveillance Framework

The envisioned use of such a probabilistic BN model is as a foundation of population surveillance of the geographical outbreak and spread of Covid-19. The proposed infrastructure for personalised Covid-19 status feedback and collecting geographical data is shown in Figure 5, and is inspired by related research of the authors' research groups (van der Heijden et al, 2013; Velikova et al, 2014).

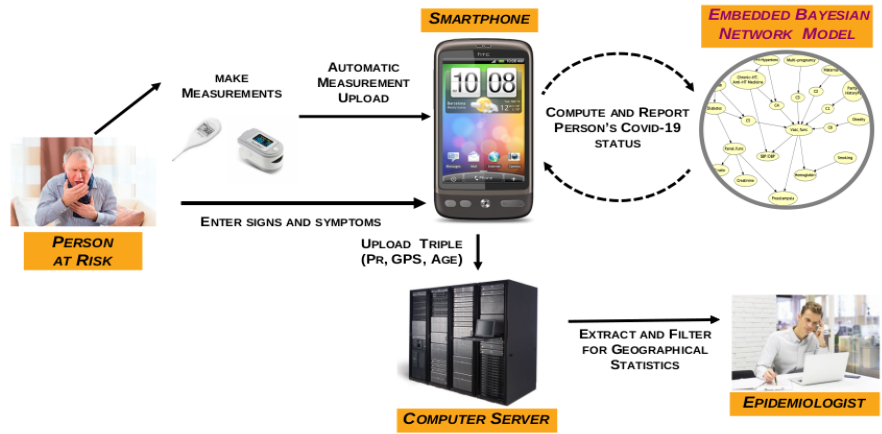

**Figure 5:** Infrastructure for personalised Covid-19 feedback and collecting geographical Covid-19 data.

As Figure 5 illustrates, the BN is embedded or integrated into an app meant to run on a person's smartphone. The presentation of the feedback is expected to be attractive and easily understood by the smartphone user with additional advice whether or not it is wise to contact a GP.

This solution operates within the CardiPro environment using the Web/PWA front-end and Agena CloudAPI (McLachlan et al, 2020). Our research group has the means now to demonstrate both the elements and the entire solution presented in Figure 5. The minimalist data transmitted to the server, even if coupled with collecting a similar anonymous symptom set as used for the Chan/Spector app, might be more palatable to people who may be concerned about privacy in both the UK and Netherlands.

In summary, in this proposed solution, it is assumed that a *citizen of a country obtains feedback about the likelihood of the presence of mild or severe Covid-19* from a smartphone app, but the main purpose of making an app with the BN embedded is *to monitor the population for detecting new outbreaks and the locations at which this occurs as early as possible.* For this purpose, it is only needed that the minimalistic data triple *(Prob. user has Covid-19, GPS-location, Age-group)* is collected centrally. The age information might be useful to get information about required protection of particular groups. In addition it might be useful to also add an app-specific unique identifier so that it is possible to follow the progress of Covid-19 in the individual (possibly until hospital admission). However, collecting only the above-mentioned data triple has the advantage of *minimal infringement of privacy*.

# Discussion

On installing CTA some personally identifiable information is always captured as a result of downloading the app from the app store, and for some apps, like the Australian and UK ones, when registering on first use (Maddocks, 2020). At a minimum, this is information which when combined with the metadata being generated makes every user and their associations identifiable. We contend that many claims regarding privacy and efficacy of CTA for COVID-19 in these papers may not be justified, and in some cases are misleading. We are not the first to identify the falsity of attempts at CTA privacy (Berke et al, 2020; Kuhn et al, 2020), nor to raise concern regarding the efficacy and applicability of CTA for COVID-19 contact tracing.

However, we are the first to consider both issues together, and as a result to demonstrate that Bluetooth CTA are not the COVID-19 panacea we all seek.

The UK and several other countries including Australia, Singapore and Germany, propose a centralised approach whereby data will be collected on smartphones and some component of that data is forwarded to a central server, enabling contact alerting and tracing of the epidemic. Some countries favour use of the solution presented by the Apple and Google Partnership, which is claimed to be a 'local' solution under development that will not breach data security and will not lead to any centralisation of data. Their proposal for privacy-safe contact tracing using Bluetooth would, they say, require explicit user consent, which is another issue that needs greater consideration. The Apple/Google solution APIs on first blush don't appear to collect personally identifiable information or user location data, and suggest a list of people you've been in contact with never leaves your phone (detected via Bluetooth LE). We are also told that people who test positive are not identified to other users, Google or Apple. That the information would only be used for contact tracing by public health authorities for Covid-19 pandemic management which in itself, like every other proposed decentralised system would necessitate communication with and storage of data in some form of central server. However, with regard to metadata and privacy, we are circumspect that Apple/Google or the various HA using their APIs will not be collecting at least part of the data being generated for secondary use purposes. It should also be noted that the Apple/Google APIs are simply an interface for HAs to expedite development of CTA solutions: they are not a CTA. APIs act as a standardised intermediary, in this case between the user interface and a data backend, both of which will still require HAs to engage software architects and developers to create. There is no guarantee that without engaging far more experienced technologists and serious reconsideration, any app the NHS develop using the Apple/Google APIs will not fare as badly as the first 24 hours of real-world testing of the NHSX CTA on the Isle of Wight (Duell, 2020). If we are to use these APIs, a better solution might be a Progressive Web App (PWA). A single PWA could be developed to be compatible with both Android and Apple architectures, and engineered to avoid the main issue seen with the NHS trial app: incompatibility with variants of the smartphone's operating system. We have already developed and demonstrated an example of this approach, called CardiPro (McLachlan et al, 2020).

> **Australia**
>
> Rather than develop their own app, the Australian Government licensed rights to rebrand the TraceTogether app developed by the Singaporean Government, and deploy it as COVIDSafe. As is common, emergency legislation was hurriedly drafted and enacted under the catchy title: *Biosecurity (Human Biosecurity Emergency)(Human Coronavirus with Pandemic Potential)(Emergency Requirements - Public Health Contact Information) Determination 2020* Act (PHCIA, 2020). While making it an offence for *a person* outside those employed by a state or federal health authority to collect, use or disclose COVID app data except for the purposes of contact tracing (Section 6(1) & (2)), this determination explicitly limits the same provision to data generated within the app or by the Commonwealth ***and*** stored on the user's mobile device. PHCIA also excludes from all provisions, privacy or otherwise, information arising from any source other than the National COVIDSafe data store (Section 6(3)). The effect of provisions of the PHCIA make it unlawful for an app user or member of the general public to decrypt, view or disseminate any data from their device, or even knowledge about data that the app collects or stores, while leaving Government organisations able to interact with this data more freely.
>
> The PHCIA contracts itself out of provisions of the *Privacy Act 1998* that may be found inconsistent under power of Section 477(5) of the *Biosecurity Act 2015*, but does not exclude itself from the operation of others, including the *Telecommunications (Interception and Access) Act 1979* (TIAA, 2018) which invokes data retention provisions on telecommunications providers, including your telephony and internet service providers, and Amazon Web Services who will be the web host of the central server, to store records of all forms of electronic communication for at least two years. The TIAA also makes metadata available without warrant to a broad range of organisations that include law enforcement, local, state and federal government bodies, the RSPCA, the Australian Navy and Border Protection Services, the Thoroughbred horse and greyhound racing associations, Workplace Safety investigators, the Clean Energy Regulator, National Measurement Institute, Building and Construction Commission, Taxi Services Commission and in some cases it has been demonstrated, private investigators (Farrell, 2016; Guy, 2015).

It can be inferred from the literature, mass media and download pages of those developing and promoting CTA, that to at least some degree they seek to create the belief that implementation of contact tracing makes containment of COVID-19 a fait accompli. Each presents a solution couched in words suggesting that, for successful eradication of COVID-19, we need only to install the CTA, and in doing so we will have identified everyone who, symptomatic or asymptomatic, might have the disease. However, this assumes the data collected by the CTA will be clean, accurate and sufficiently complete, and will fully support their containment efforts which, despite best intentions, is extremely unlikely (Senga et al, 2017). We accept that solutions operating at the front end of contact tracing, like the CTA, will produce more data. More contact information will require time-consuming and labour-intensive follow-up, and consumption of considerable resources in order to identify and weed out the true cases from the spurious chatter (Senga et al, 2017). But it should be noted that previous work has failed to consider:

    a) the effect of people simply leaving their smartphone at home, or in the car;

    b) how to effectively deal with people who might have two or more devices; or

    c) how to identify the owners of prepaid devices that in some countries can be registered without identification, or anonymously.

    d) the effect of a CTA user coming into close physical contact with others who eschew, or cannot afford, smartphones (all previous work assumed that the adjectives *pervasive* and *ubiquitous* meant *complete coverage*).

We believe that care should be taken when deploying CTA in any community. Not just because of privacy or consent issues. But rather, to ensure that even the most suggestible member of our community does not become complacent and assume that CTA operates, as claims like those provided with the Australian Government COVIDSafe app would seem to suggest, as an invisible shield making us and our families impenetrable or immune to the disease.

Even if all potential privacy issues were resolved, the decision to install and register the CTA in most western countries would remain voluntary. This raises the question: How can high-level uptake of the CTA be assured? To answer this question we propose that at least three related matters must be considered: (i) public compliance with existing social distancing measures; (ii) media narrative of CTA; and, (iii) ongoing changes in peoples' subjective estimate of severity and susceptibility to the virus.

Opinion polls in recent weeks are finding that the majority in each country are in favour of existing social distancing measures, irrespective of how strictly they are maintained and for how long they remain (Ipsos-mori, 2020a). When compared to other countries, people in the UK are displaying a higher degree of support for continued social distancing. Similarly, the UK population has shown overwhelming support for lockdown measures, again irrespective of severity and duration of the lockdown. It may be that this public acceptance will extend to other State-operated measures including the suggested *test, track and trace* strategy that includes the CTA, as it is being promoted as a way to end the lockdown and reduce the possibility for additional and more severe lockdown measures. It is possible that, irrespective of how privacy-invading CTA methods may be, or the potential negative impact that third-party use of metadata resulting from individual engagement with the app, the public may accept these impositions in return for the benefits of a lifted lockdown and lighter social distancing measures. Certainly, the polling conducted between the 10$^{th}$ to 13$^{th}$ of April 2020 in the UK suggests this holds true, with 65% showing support for the CTA (Ipsos-mori, 2020b). However, public opinion elsewhere is somewhat mixed.

> **United Kingdom**
> While drawing significant criticism, the UK National Health Service (NHS) has rejected the Apple/Google APIs and decentralised model, expressly favouring a centralised approach that they say will allow for collection of more granular data and broader analysis to study and track the pandemic (Hamilton, 2020). The key difference to be noted between the NHS approach and all others is upfront acknowledgement of the intention to maintain this central collection of data while also making substantial claims regarding the privacy strength of the userland app and ethics of their approach. Unlike descriptions of all other claimed privacy-preserving apps seen in the COVID-19 literature, and in stark contrast to the Australian approach of denying the public any real knowledge of the data being collected and transmitted by their device (PHCIA, 2020), the NHS are making encouraging noises regards allowing researchers, security analysts and the general public access to the source code, to see behind the curtain and verify what data the app is collecting and transmitting (Gould & Lewis, 2020). Unlike any other, and if taken on face value, this could allow UK citizens to consider that data's existence and potential uses when deciding whether to download and activate the app on our personal devices.

In other countries the trade-off is not the same: the protection of privacy outweighs relaxed social distancing through use of a CTA. For example: (i) in France, where 53% of respondents are opposed to the CTA (Hughes Hubbard, 2020); and, (ii) the US, where 50% of respondents are opposed to the CTA, (Kirzinger et al, 2020). The US poll also showed opinion somewhat changes when benefits such as going back to work are more prominently presented, in which case 66% would agree to download the CTA. However, from 64% of the US total sample 17% indicated that a CTA would make them feel less safe, while 47% said the CTA would make no difference to their feelings of safety at all. The current media narrative and an individual's subjective estimates of severity and susceptibility are two broad factors that, whilst not independent of each other, account for the observed differences in opinion and behaviour both between countries and over time (Abeysinghe & White, 2011; Leppin & Aro, 2009; Slovic, 2000; Wagner-Egger et al, 2011; Wheaton et al 2011).

The contentious issue of privacy presents as a far more salient and palatable target than dealing with the overwhelming lack of evidence for efficacy. On these issues there are now several open letters from scientists that are being communicated to the general public. While the sustained focus on data privacy concerns remains strong in the mainstream media, this negative issue will dominate public understanding of CTAs and significantly restrain uptake. If the narrative can be drawn towards the potential benefits for everyone that come from a general loosening of restrictions to open schools and workplaces, then the success we have seen in compliance with the current lockdown may allow these people to accept the trade-off and come out in favour of the CTA. Naturally, this won't be isolated from individual's estimates of severity and susceptibility to the virus, and by extension, for those close to them. But if there is a sufficiently strong belief that severity and susceptibility is high in those close to oneself, even if the severity and/or susceptibility is low for themselves, then, just as we have seen with compliance to lockdown measures, compliance with State messaging on voluntarily using a CTA may also be high.

> **The data being collected**
>
> Drawn from many of the cited papers in this work, most apps will collect and transmit some subset of the following data fields:
> - MAC address of your device's Bluetooth or Wi-Fi chip
> - Your Phone number (or IMEI number if the device does not easily report the subscriber phone number)
> - The MAC address of other people your phone sees (Bluetooth handshakes with everything it sees that is also Bluetooth, even when it doesn't know the device and has never been paired with it)
> - The time, date and in some cases, location data from your GPS for each new interaction with another in-range device (accurate to about 15 meters). A new interaction is when your device sees another device move into its broadcast area. Note that in a corporate office the app might see the device of someone in the next room move into and out of range tens or hundreds of times over the course of a working day.
> - The Bluetooth or device name of the smartphone that is running the app, and every other Bluetooth device that crosses into its broadcast range. This last point can more easily enable re-identification as people often name their smartphone 'Tim's iPhone' or similar.

Many proposed solutions, even the Google/Apple collaboration, focus very heavily on privacy and app distribution and make almost no mention regarding accuracy. Despite best intentions, the levels of inaccuracy that arise in any data recording mean that any contact tracing, manual or digital, will always be incomplete (Senga et al, 2017). Even when we have a significant proportion that do comply with contact tracing, we often still have poor data arising out of the methods employed to collect the data. The normal inaccuracies that occur in data recording and data entry are amplified with contact tracing because some people simply don't want to be traced, while others have limited socio-cultural understanding for why we are wanting to trace them (Senga et al, 2017). Contact tracing represents an expensive win/lose situation. A very small group of university researchers and technology companies receive a large funding boost to develop the CTA and deal with the data that it collects, and a large number of people involved in manual contact tracing win jobs. However, the overall community suffers more significant risks, and losses, when they choose to re-engage in normal behaviours under the false sense of hope that most CTA are promoted as giving, and risk becoming infected and infecting their family, potentially leading to death.

We are sceptical that any standalone contact tracing approach, manual or automated, could contain a high-prevalence highly contagious disease like COVID-19. This is primarily because the CTA acts retrospectively. It advises the user they were previously in close contact with an infected, and in the case of COVID-19, this advice often comes only after they have already begun asymptomatically shedding the disease. The primary (third) solution we propose integrates the retrospective CTA with symptom tracking and a BN, providing the user with a prospective view of the probability that they may have contracted COVID-19. In this way we increase CTA utility for users. We believe that with increased utility uptake may be improved, as is the opportunity to collect useful data and identify actionable clinical knowledge to improve the response in future disease outbreaks. Solutions like the one proposed here can have a very beneficial effect on containing the spread of infection.

One final point that returns to the issue of consent that was raised earlier. For many proposing CTA, the idea of using an app instead of just network tracing via the cellular network or other means is not as much about Bluetooth being more accurate, it is about the idea of claiming to have informed consent: that by downloading the app and clicking through a privacy agreement they have received 'informed consent' to access and monitor an individual through their device. In studies evaluating the impact and effect of privacy policies and user agreements it was found that 54% are written in language unapproachable by most people (Jensen et al, 2004), 40% of participants do not even recall seeing the agreement while clicking through to install the app (Good et al, 2005), and only 0.24% of more than 55,000 actually clicked or scrolled to view the policy (Jensen et al, 2004). Most users have no idea what they have agreed to, and given that organisations change their policies and agreements regularly, whether the current version of the agreement is consistent with that which the media may have discussed when the CTA was being rolled out. Given these findings, is it ethical to consider that when users install and register the CTA, the inclusion of a long privacy policy and user agreement that potentially more than half of the population will be unable to comprehend constitutes *informed consent*?

# Conclusion

In writing this paper we reviewed a large collection of topical works on COVID-19. Most works were recent preprints proposing CTA solutions for containment of the disease, while others presented the latest research and evaluation of the spread of the disease in our communities. While there is a focus in the literature on two issues, privacy and efficacy, the current media narrative for CTA in many countries strongly emphasises perceived privacy risks, and in the UK especially, the risks some attach to the NHS decision to eschew the presumed leading solution: the Apple/Google collaboration.

We also sought to simulate the operation of CTA, and the results of our calculations appear largely in agreement with those of other groups just published. Introduction of a new CTA alone would not contain the disease, and the best-case sweet spot for uptake is beyond that which could conceivably be achieved. However, by providing people with an understanding not just retrospectively for whether they have been in contact with an infected previously, but also using a Bayesian approach to proactively provide the probability that they might have the disease, we can increase the CTAs utility to users while potentially improving the uptake and knowledge to be learned from use of the app. When combined with an effective communication strategy and sensible social distancing, we believe solutions like the one proposed here can have a very beneficial effect on containing the spread of this pandemic and reducing the need for draconian lockdown procedures.

# Acknowledgement

The authors acknowledge support from the EPSRC under project EP/P009964/1: PAMBAYESIAN: Patient Managed decision-support using Bayes Networks